%% file: main.tex
\documentclass{article}

\usepackage{PRIMEarxiv}

\usepackage{chrkroer}
\usepackage{pifont}
\usepackage{algorithm}
\usepackage[noend]{algpseudocode}
\usepackage{subcaption}
\usepackage[normalem]{ulem}
\usepackage{natbib}
\usepackage[capitalise]{cleveref}
\usepackage{amsfonts}   

\newcommand{\horizon}{\mathcal{T}}

\newcommand{\ck}[1]{{\color{red}** CK: #1 **}}

\newif\iflong
\longfalse

\title{Fair Notification Optimization: An Auction Approach}

\author{
  Christian Kroer \\
  Meta and Columbia University \\
USA \\ 
   \And
   Deeksha Sinha, Xuan Zhang, Shiwen Cheng, Ziyu Zhou \\
   Meta \\ 
   USA
}

\begin{document}
\maketitle

\fancyhead[LO]{Kroer, Sinha, Zhang, et al.}

\begin{abstract}
    Notifications are important for the user experience in mobile apps and can influence their engagement. However, too many notifications can be disruptive for users. A typical mobile app usually has several types of notification, managed by distinct teams with objectives that are possibly conflicting with each other, or even with the overall platform objective. Therefore, there is a need for careful curation of notifications sent to users of these different types. In this work, we study a novel centralized approach for notification optimization, where we view the opportunities to send user notifications as items and types of notifications as buyers in an auction market. Furthermore, the auction setup is unique, and the platform has the ability to \emph{subsidize} the bids from the notification types. 

    Using tools from fair division, we study the application of competitive equilibrium for addressing this problem. We show that an Eisenberg-Gale-style convex program allows us to find an allocation that is fair to all notification types in hindsight. 
    Using the dual of the formulation, we present an online algorithm that allocates notifications via first-price auctions using a pacing-multiplier approach.
    Secondly, we introduce an approach based on second-price auctions and pacing, which has the benefit of working well with existing advertising systems built for second-price auctions.

    Through an A/B test in production, we show that the second price-based auction system improves over a decentralized notification optimization system, leading to its launch in production for some Instagram notifications. Further, through simulations on Instagram notification data and a subsequent production A/B test, we compare the outcomes of first-price and second-price auctions and show that the former has more stable pacing multipliers.
\end{abstract}


\maketitle

\section{Introduction}

Notifications are an important part of many mobile application (app) experiences. They inform users about promotions, new emails and messages, relevant activity in their social network, reminders about events such as birthdays, updates to threads they follow, etc. By serving as a communication channel for important information, notifications can influence user engagement on apps. Further, push / email notifications provide a way to communicate to users when they are not actively using the app. Though on the flip side,  notifications can  be a source of distraction and too many notifications can make users feel spammed. This can lead to users becoming habituated to ignoring notifications or turning off notifications, cutting off the possibility of any future communication through this channel. Thus, there is a need for careful curation of the notifications sent to users, sending only those that provide the most value to them.

\begin{figure}[t]
    \centering
    \begin{subfigure}{.44\linewidth}
        \centering
        \includegraphics[width=\textwidth]{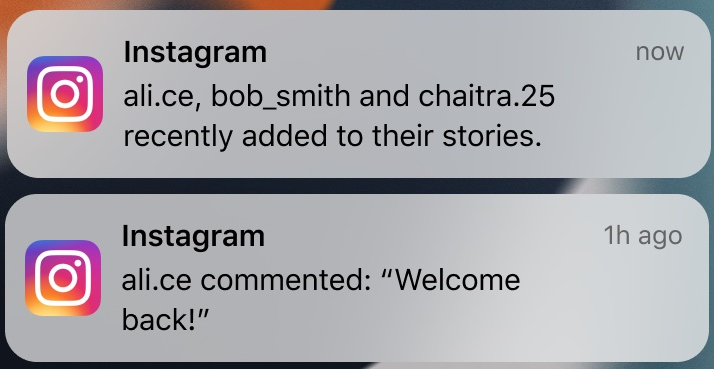}
    \end{subfigure}%
    \hspace{.02\linewidth}
    \begin{subfigure}{.5\linewidth}
        \centering
        \includegraphics[width=\textwidth]{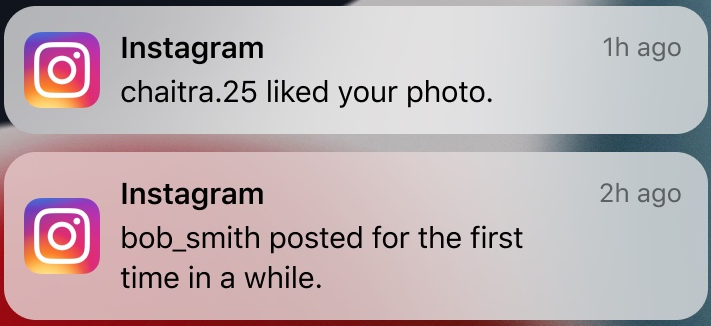}
    \end{subfigure}
    \vspace{-1em}
    \caption{Sample of push notifications\vspace{-2em}} 
    \label{fig:push-notif-sample}
\end{figure}

For the purpose of this curation, we need to understand the goals behind sending notifications. Often there are several types of notifications even within a single mobile app. For example, in a social media app, there can be notifications informing users about engagement (e.g., likes and comments) on their posts, new posts made by their friends, events such as birthdays, etc. For context, there are more than 550 types of notification in the Instagram app (see Figure \ref{fig:push-notif-sample} for some examples). These notification types are managed by individual product teams, each having different goals. Typically, each team is responsible for choosing an appropriate set of notifications of that type to send to users. But this operating model has multiple shortcomings. 

Different teams can have goals that are not aligned with each other or with the overall platform\footnote{We use the terms mobile app and platform interchangeably in this text.} objectives. Thus, optimization of notifications within the silo of each team can lead to incoherent communication for the users. In addition, notifications from different teams may reach the same set of users and hence compete with each other for the users' attention. Furthermore, if one team sends excessive notifications to a user, making the user unresponsive to future notification, all other teams are also adversely impacted. Another limitation of this decentralized approach is that the best notification of one type might not be among the best when all notification types are considered. This leads to suboptimal notifications being sent to users. Thus, it is important for these teams to work cohesively to identify the best set of notifications across all types to send to users. This set of notifications should not only take into account the goals of each team but also the overall goals of the app. 

More generally, this can be viewed as a problem of fairly sharing users' attention across multiple parties with potentially different goals. This also occurs on the personalized landing pages of many websites (or mobile apps) such as Spotify and Amazon. On Amazon's homepage, there are multiple different sections such as ``Top Deals", ``New Year Sale", ``Best Sellers in Books", ``Amazon Live''. These sections are likely maintained and curated by different teams within the company, each responsible for the sale of a different category of products. Thus, for each user, these teams have to decide cohesively what are the best sections to show on the personalized homepage. For the rest of the paper, we focus on the notification application, although the model and insights are broadly applicable to these other settings.

In this work, we present an auction-based system for optimizing over notifications. This system incorporates the goals of each notification type team and the platform. 
Using tools from fair division, we study the application of competitive equilibrium for ensuring fair allocation of notification sending opportunities. We show that an Eisenberg-Gale-style convex program allows us to find an allocation that is fair to all notification types in hindsight, due to the fact that the formulation optimizes the \emph{geometric mean} of the utilities of notification types and the platform. Using the dual of the formulation, we present an online algorithm that can be used to fairly allocate notification opportunities over time, by allocating each opportunity via first-price auctions combined with pacing.

We describe a production implementation and relevant design choices for this auction system. The production design choices highlight both the novel aspects of this auction system and the elements based on auction systems already in use for other applications in the company (particularly the pacing algorithm and auction strategy). The latter enable re-use of existing production infrastructure. Through an A/B test, we validate the gains achieved from the centralized auction system over the existing decentralized optimization system which lead to its launch in production.

Further, we compare, through simulation on real Instagram notification data and a subsequent production A/B test, the outcome of the proposed pacing algorithm with the outcomes of a first-price and a second-price auction using the production pacing algorithm. We observe that the two first-price auctions have similar outcomes albeit their pacing approaches differ. Further, we observe that first-price auction has more stable pacing multipliers laying ground for further improvements in the production system.

\subsection{Related Work}

Several studies have established the value of notifications for engaging users on websites and mobile apps \citep{glay2019real, 7756930, wohllebe2020consumer, wohllebe2021mobile}. At the same time, it has been established that excessive notifications can cause the opposite effect \citep{7756930, wohllebe2020consumer, wohllebe2021mobile, mehrotra2016my}. This need for balance and communicating the most important information with few notifications has driven research on how to optimally send notifications to users. This stream of research has been tested and deployed in production systems in companies such as Twitter \citep{o2022should}, LinkedIn \citep{gupta2016email, gupta2017optimizing}, Pinterest  \citep{zhao2018notification}  and Duolingo \citep{yancey2020sleeping}.

\citet{tan2016exploration} discuss metrics for evaluating quality of push notifications. Several types of techniques have been proposed in the literature for the notification optimization task. \citet{gupta2016email, gupta2017optimizing, gao2018near} formulate a problem to minimize the number of notifications sent (email / in-app/ push) with constraints on the resulting user sessions or platform wide events. 
Here, the optimization problems are solved using a linear program for the offline version and a primal-dual approach for the online problem.

Several other techniques have also been proposed for the notification optimization problem. Given that long term user engagement is likely driven from a sequence of notifications, instead of one notification, there have been works on modeling the notification decision-making as a Markov Decision process and subsequently using reinforcement learning techniques for optimization \citep{yuan2022offline, o2022should, prabhakar2022multi}. 
\citet{yuan2019state} propose a survival analysis based framework for modeling the effectiveness of notifications. \citet{zhao2018notification} propose a machine learning approach to decide notification volume for each user. \citet{yancey2020sleeping} propose a multi-armed bandit approach for notification optimization. 
\citet{yue2022learning} propose a ranking solution to decide which notification to send to users. 

Most of the above works optimize for a single objective function. \citet{prabhakar2022multi} consider optimizing for multiple objectives by scalarizing to a single objective using a weighted combination. In contrast to these, our work considers multiple objectives of different notification teams and the overall platform goals and seeks to distribute the notification resources fairly to optimize these objectives. We present an auction based solution for this optimization problem. 
We discuss guarantees of this auction system and present appropriate pacing algorithms. Importantly, the auction based solution provides fairness guarantees for the different notification types and the platform.
An earlier system-focused paper~\citep{zhu2021nonmonetized} also discusses a pacing-based second-price notifications system, and our paper builds on that work. We propose to study both first-price and second-price auctions, and develops the first algorithmic and fairness guarantees for such systems. 

The Competitive Equilibrium from Equal Incomes (CEEI) approach that we base our first-price auctions approach on was introduced by \citet{varian1974equity}, where he showed that the CEEI mechanism has many desirable properties such as envy-freeness, proportionality, and Pareto optimality. 
Since then, the CEEI mechanism, or variants thereof, has been applied to a number of real-world problems such as matching course seats to students~\citep{budish2011combinatorial}, allocating food to food banks~\citep{aleksandrov2015online,sinclair2021sequential}, fair resource allocation for contracted ads~\citep{bateni2022fair}, and fair allocation of workforce to different sections \citep{allouah2022robust}.
Moreover, competitive equilibria in Fisher markets are known to be related to budget-management systems for first-price auctions in internet advertising settings~\citep{borgs2007dynamics,conitzer2022pacing}.
In all of the above work, it is assumed that the utility of each buyer depends only on the items that they receive. Our competitive equilibrium model is novel in that it introduces an auxiliary buyer, the platform, which does not receive items, but instead cares about how items are matched to the other buyers, i.e. the notification types.
The problem of computing Fisher market equilibria has received extensive attention in the computer science literature, with both combinatorial and first-order algorithms~\citep{shmyrev2009algorithm,devanur2008market,birnbaum2011distributed,gao2020first}.
Second-price auctions have also been studied extensively in the context of budget-management systems in internet advertising~\citep{balseiro2017budget,balseiro2019learning,conitzer2022multiplicative}.
Most related to our work is \citet{conitzer2022multiplicative,borgs2007dynamics}, who show a connection between Fisher markets with \emph{supply-aware} buyers and the solution to budget-management problems in second-price auction markets.

\subsection{Notification System Overview}

In this subsection, we give an overview of the notification system on Instagram.
Notification generation can be triggered by a variety of events depending on the notification type. For example, a notification can be generated for a particular user due to relevant activity by another user (such as a comment, or a relevant post) or it can be triggered at fixed times such as to summarize recent activity in the user's network. After a notification is generated, the sending optimization layer makes a real time decision if the notification should be sent to the user or not. If the decision is in favor of sending the notification, then the notification is sent to the user immediately. Otherwise, the notification is dropped. In summary, the generation layer is responsible for what and when to generate notifications for a user; and the sending layer is responsible for deciding which of the generated notifications to send to the user. In this work, our focus is on optimizing the sending layer, while taking the notification generation process as input. 

\subsubsection{Decentralized sending optimization system}

The notification sending system before the introduction of the auction system was based on a decentralized optimization framework. In this system, each notification type individually optimized for which notifications to send. Machine learning models are trained on each notification type to quantify the value of a notification with respect to the relevant metrics for that notification type. For example, the models can be trained to predict the probability of the notification being clicked on by the user or the notification leading to the user using the app or a particular feature on the app that day. Each notification type team tunes thresholds for these model scores, taking into account the desired sending volume of that notification type and the users' response to the volume and quality of notifications. A notification is sent to the user if the notification's score is greater than the threshold.

\subsubsection{Auction based sending optimization system}

In this work, we study a centralized mechanism where a central agency, instead of individual notification teams, decides the set of notifications to send. The idea of using an auction system in non-monetized markets in Meta's line of businesses beyond the ads space was first proposed by \citet{zhu2021nonmonetized}. In this system, different notification type teams are given a budget of faux currency which they use to bid on the opportunity to send a notification to a user. 
The auction system consists of the following key components.

\paragraph{Notification type bidders} Each notification type has a model which estimates the value of generated notifications, which are related to the goals of the team responsible for the notification type. These valuations guide the notification type bidders on the amount to bid on the generated notifications.

\paragraph{Platform bidder} The platform also has a model which estimates the value of generated notifications based on platform metrics, which could be different from those of the notification types. For example, a notification type might prioritize engagement on a specific product whereas the platform is interested in broader platform level engagement metrics. The platform bidder similarly decides their bids based on their valuations of generated notifications.

\paragraph{System bidder} The system bidder serves as a personalized time-varying threshold for controlling the number and quality of notifications sent to users. A notification is sent only if its combined bids from the notification type and the platform exceeds the system bid. Thus, the system bids can be viewed as reserve prices on the opportunity to send notifications to users. 

\paragraph{Budget} Each notification type, as well as the platform, is assigned a faux currency budget that they can use to bid on notifications. These budgets serve as a lever to control the resource distribution among the (possibly competing) goals of different notification types and the platform.

\paragraph{Pacing} The pacing system ensures that the budget is spent uniformly over time and in accordance with the expected amount of future notification opportunities. Furthermore, the pacing system helps make the valuations from the different notification types and the platform comparable by bringing them to the same budget unit. 
The system maintains a pacing multiplier for every notification type and the platform, and adapts it according to their overall budget and spending rate. The bids are computed as follows.
\begin{small}
\begin{align*}
    \text{notif. type bid} &= \text{notif.~type valuation}  \times \  \text{notif.~type pacing multiplier}  \\
    \text{platform bid} &= \text{platform valuation} \times \  \text{platform  pacing multiplier} 
\end{align*}
\end{small}
The system also has the ability to perform pacing on the system bid. This lever is typically used to ensure different sending rates for users in different segments. The sending rate is defined as the ratio of notifications sent to notifications generated.

\paragraph{Auction instance} An auction is run as soon as a notification is generated. Since it is rare that two or more notifications are generated for a user at the same instant, most auctions consist of exactly one notification type bidder.

\input{model_notation.tex}

\input{first-price-auctions}
\input{second-price-auctions}

\section{Online Algorithm for {nFPPE}}
\label{sec:algorithms}

We start the section by deriving the dual of \eqref{eq:mnw}, which will be used to develop the online algorithm.

\begin{proposition} \label{prop:dual}
The dual of \cref{eq:mnw} is the following:
\begin{small}
\begin{align} 
\begin{array}{rlll}
    \displaystyle \min_{p, \lambda, \beta \ge 0} & \displaystyle \sum_{j\in [m]} s_j p_j + \sum_{t\in [\horizon]} \lambda^t - \sum_{i\in [n]} B_i \log \beta_i - B_p\log\beta_p \\[6mm]
    \text{s.t.} & \displaystyle p_{j(t)} \ge \beta_{i(t)} v^t + \beta_p v_p^t -\lambda^t \;\; \forall t\in [\horizon].
\end{array} \label{eq:mnw dual}
\end{align}
\end{small}
\end{proposition}

Since we want to perform the allocation online, it will be useful to reformulate this problem in order to get rid of the $\lambda^t$ variables. After reformulating, we get the following dual problem, where $[\cdot]^+ \coloneqq \max(\cdot, 0)$ denotes thresholding at zero.
\begin{footnotesize}
\begin{align}
    \min_{p, \beta \geq 0}
    & \sum_{t=1}^{\horizon} \left[\beta_{i(t)} v^t + \beta_p v_p^t - p_{j(t)} \right]^+
    + \sum_{j=1}^m s_j p_j
    - \sum_{i=1}^n B_i \log \beta_i   
    - B_p \log \beta_p
    \label{eq:mnw pairs dual}
\end{align}
\end{footnotesize}

\begin{algorithm}[t]
\begin{algorithmic}[1]
    \State{\textbf{Input:} pacing multipliers $\{\beta^0_i\}_{i\in [n]}$ and $\beta^0_p$, prices $\{p_j\}_{j\in [m]}$}
    \State{initialize $u_i=0$ for all $i\in [n]$ and initialize $u_p=0$} 
    \For{time $t=1, 2, 3, \cdots, \horizon$}
    \State{a notification of type $\tau_{i(t)}$ is generated for user $u_{j(t)}$ with valuation $v^t$ for the notification type and  $v_p^t$ for the platform}
    \If{$\beta^{t-1}_{i(t)} v^t + \beta_p^{t-1} v^t_p \ge p_{j(t)}$}
    \State{set $x^t=1$}
    \State{update notification type utility: $u_{i(t)}\leftarrow u_{i(t)} + v^t$}
    \State{update platform utility: $u_p \leftarrow u_p + v^t_p$}
    \EndIf
    \For{\textbf{each} notification type $\tau_i$}
    \State{update pacing multipliers: $\beta_i^t \leftarrow \min\left(\frac{B_i}{ (u_i/t)}, \frac{B_i}{\underline u_i}\right) $}
    \EndFor
    \State{update platform multiplier $\beta_p^t \leftarrow \min\left(\frac{B_p}{(u_p/t)}, \frac{B_i}{ \underline u_p}\right)$}
    \EndFor
    \State{\textbf{Output}: allocation vector $\{x^t\}_{t\in [\horizon]}$}
\end{algorithmic}
\caption{} \label{algo:online}
\end{algorithm}

Our algorithm (outlined in \cref{algo:online}) assumes that a set of per-user prices $\{p_j\}_j$ is given, and focuses on learning the remaining variables, which are the pacing multipliers $\beta$.
The user prices $\{p_j\}_j$ are assumed to come from some black box, for example, the price that would have yielded the right amount of notifications yesterday, or a machine learning model. 
The reason we do not perform online learning of the user prices is that most users only have a relatively small number of potential notifications generated each day, and thus we do not expect to be able to learn those prices online.

Then, for a fixed set of prices $p$ for each user, we can use online learning to solve \eqref{eq:mnw pairs dual}:

\begin{footnotesize}
\begin{align}
    \min_{\beta \geq 0}
    & \sum_{t=1}^{\horizon} \left[\beta_{i(t)} v^t + \beta_p v_p^t - p_{j(t)} \right]^+
    + \sum_{j=1}^m s_j p_j
    - \sum_{i=1}^n B_i \log \beta_i   
    - B_p \log \beta_p
    \label{eq:mnw pairs dual beta only}
\end{align}
\end{footnotesize}

\smallskip\textbf{The underlying solution}. 

Now suppose that there is an underlying probability distribution $\mu\in L_+^\infty$ over possible valuation matrix pairs $\theta = (v, v_p)$. Suppose furthermore that the support of $\mu$ is contained in $[0,1]^{n\times m} \times [0,1]^{n\times m}$. Note that this model is more general than our notifications setting in the sense that this model allows multiple interested notification types per time period (where a single sampled $(v,v_p)$ is analogous to a time period).
Suppose we treat each pair $\theta$ as an item with supply $s(\theta)$. 
In that case, we get an infinite-dimensional Fisher market whose market equilibria are described by the following convex program:
\begin{equation}
\begin{aligned}
    \min_{p\geq 0,\beta \geq 0} 
    & \sum_{j=1}^m \mathbb E_{(v_{ij},v_{pij}) \sim \mu} \left[ \max_{i} \left(\beta_i v_{ij} + \beta_p v_{pij} - p_j \right) \right]^+ \\&
    + \sum_{j=1}^m s_j p_j
    - \sum_{i=1}^n B_i \log \beta_i   
    - B_p \log \beta_p. 
\end{aligned}
\label{eq:mnw pairs inf}
\end{equation}
We call this the \emph{underlying market equilibrium}.
This is analogous to the infinite-dimensional Fisher market model of \citet{gao2022infinite}, except that we include the platform buyer $p$, and the per-user supplies $s_j$.
Their results about the relationship between solutions to \cref{eq:mnw pairs inf} and market equilibria easily extend to our setting, since the fact that valuations are contained in $[0,1]$ means that all measurability and integrability conditions are satisfied.
Thus, there is a vector $u^* \in \mathbb R_+^n$ describing the utilities achieved by each buyer $i$ in the underlying equilibrium. 

\smallskip\textbf{PACE algorithm}.
Now we show that if we are given the user prices $p^* \in \mathbb R_+^m$ associated to the supply constraint for each user, then we can learn the pacing multipliers $\beta_i,\beta_p$ from \cref{eq:mnw pairs inf} in an online fashion.
Our algorithm is an extension of the PACE (Pace According to Current Estimated utilities) algorithm of \citet{gao2021online}. The algorithm is given in Algorithm \ref{algo:online} for the case where only a single notification type has a non-zero valuation at each time $t$ (thus corresponding to our nFPPE setting).

\begin{theorem}\label{thm:pace}
    PACE generates pacing multipliers $\beta^t$ (where we include the platform multiplier $\beta_p^t$ in this vector) such that 
    \[
        \bbE \|\beta^t - \beta^*\|^2 \leq \frac{(6+\log t)G^2}{t \min_{i\in \{[n],p\}}\underline u_i^2},
    \]
    where $G^2 = \max_{i\in \{1,\ldots,n\}} \bbE_{(v_i,v_p)\sim \mu}\big[\big(\sqrt{v_i} + \sqrt{v_p}\big)^2\big]$, and $\underline u_i$ is the proportional utility of $i$.
    \label{thm:items online pace}
\end{theorem}

\cref{thm:items online pace} shows that pacing multipliers from learning nFPPE online converge in mean-square. Thus, we can expect that an nFPPE-based system will have stable pacing multipliers, whereas for nSPPE we have no such guarantee. Indeed, we do not expect them to be stable, based on the connection to the PPAD-hard SPPE problem.

Note that the guarantee is under the assumption that user prices are given as inputs. However, in practice, it is hard to correctly estimate user prices as user activities on the platform (and thus the amount of notifications generated) are often spiky. Thus, additional heuristics are needed to learn and update user prices. We discuss more about this in Section \ref{sec:heuristics}.

\section{Experiments}

\subsection{Launch of SP in Production}
\label{sec: exp_sp_launch}

To evaluate the performance of the auction system on real data and in the production environment, we ran an A/B test to compare the auction system (\textit{test group}) with the production system at the time (\textit{control group}) for four push notification types at Instagram.

The control group was a decentralized optimization system where notifications of each type were separately optimized. A generated notification was sent to the user if its utility exceeded the notification type specific threshold. These thresholds had been tuned, separately for each notification type, to achieve a desired volume of notifications sent and click-through rate.

The test group was the auction system with \textit{budget spent based pacing}. This pacing mechanism is based on the pacing mechanism used for other auction applications within the company. The pacing multiplier is defined as a function of the ratio of the notification budget spent to the expected budget spend\footnote{Exact functional form concealed due to confidentiality concerns.}. The expected budget spend is calculated based on a uniform budget spending rate.

Second price auctions with budget spent based pacing were chosen instead of first price auctions because of two main reasons. Firstly, second price auctions are already used in the company in other domains such as ads, leading to general familiarity with the concept. Secondly, there already exists production infrastructure related to budget spent based pacing, easing its use in new domains. 

The budget for each notification type and the system bidder (which can be thought of the price) were configured through simulations and prior A/B tests. The system bidder was configured to be the fifth-highest bid of all notifications in the past three days for the user and the budgets were chosen as shown in Table

This A/B test was run from May 10, 2022 to June 11, 2022. Each user in the experiment was assigned to either the test or the control group. For a user in a particular group, all the generated notifications went through the corresponding optimization system. Both the experiment groups consisted of 77.4 million users. The four notification types considered in the A/B test were Comment Subscribed, Feed Suite Organic Campaign,  Like, and Story Daily Digest. Table \ref{tab:notif_desc} in Appendix \ref{sec:app_tables} contains the description of each of these notification types and Figure \ref{fig:push-notif-sample} contains a sample of each notification type.

The A/B test showed positive results in favor of the auction system (\Cref{tab:ab_test_results}). 
The auction system increased two core platform metrics of Instagram (metric name and definition concealed due to confidentiality concerns). Further, the click-through rate on notifications increased. Simultaneously, there was a decrease in the number of notifications sent. Further, there was no negative impact on the reachability of users due to excessive notification sending. Typically, if users get more notifications than desired then they turn off notifications which impacts reachability through push notifications. These results show that the auction system increased user engagement by sending fewer, but better quality, notifications.

\begin{table}[t]
    \centering
    \begin{tabular}{l|rr}
        \bf Metric & \bf Change (1) & \bf Change (2)   \\
        \hline
        Click-through rate & 0.42\% & Neutral \\
        No. of notifications sent* & -0.495\% & 0.28\% \\
        User reachability  & Neutral & Neutral\\
        Core platform metric 1 &  +0.057\% & Neutral\\
        Core platform metric 2 & +0.064\% & Neutral\\
    \end{tabular}
    \caption{Metric improvements: (1) from a decentralized system to an auction system; and (2) from second price auction to first price auction, both with budget spent based pacing, in an A/B test. *across all Instagram notification types \vspace{-2em}} 
    \label{tab:ab_test_results}
\end{table}

Given the strong performance of the auction system in the A/B test, it was launched in production for these four notification types on Instagram in July 2022. This launch impacts about 2.9B (resp.~13.1B) daily sent (resp.~generated) notifications. More notification types have been added to the auction system since then.

\subsection{Dataset}

For simulations, we consider a dataset with all the generated notifications for a subset of users across the four notification types mentioned above. In addition, the faux currency budget $B_i$ for each notification type is available (see Table \ref{tab:notif_desc} in Appendix \ref{sec:app_tables}).

In terms of users' capacity for notifications, the rule of thumb we consider is that no more than five notifications should be sent every three days across these four notification types. Thus, for our experiments, we always choose a three-day window and assume the time horizon $\horizon$ is the number of notifications generated during the three-day period and let $s_j=5$ for every user $u_j$.

Each row of the dataset represents the notification event at one time period, which includes (1) notification type that generates the notification; (2) user for whom the notification is generated; (3) the utility of the generated notification for the notification type; (4) timestamp. Note that the platform utility is not currently available in production. Therefore, we assume that $v^t_p=0$ for all time $t$.

\subsection{Algorithm \ref{algo:online} Heuristics} \label{sec:heuristics}

In this subsection, we describe how to obtain the inputs for Algorithm \ref{algo:online} and additional steps required to maintain these parameters. For the input, we use historical data -- in particular, the three days immediately before the three-day window for the experiment data. 

\subsubsection{Price Learning} 

Although one can learn the price by solving the optimization problem in \eqref{eq:mnw} and gather the optimal dual solution, this is essentially impossible, or at best extremely time-consuming, due to the size of the user set. However, note that the price of each user $u_j$ can be obtained by
\begin{small}
\begin{equation*}
    p_j \coloneqq \begin{cases} 
        \textup{$5$-th largest value in } \{\beta_{i(t)} v^t + \beta_p v^t_p: t\in \horizon_j\} & \textup{if } |\horizon_j|> 5; \\ 
        0 & \textup{otherwise.} \end{cases}
\end{equation*}
\end{small}

Hence, we may instead estimate the pacing multipliers $\{\beta_i\}_{i\in [n]}$ and $\beta_p$, and then compute the prices using these estimates. Pacing multipliers are much easier to estimate as they can be computed by solving a smaller optimization problem where only a random sample of the users is included. The optimization problem is solved by the Hypatia package in Julia \citep{coey2022solving}. 

\subsubsection{Price Updates}

We observe that for many users, their prices are zero because fewer than five notifications were generated during the three-day window used for price learning, but then during the three-day simulation window, they post something that results in the generation of a significant amount of ``like'' notifications. This is not surprising given that many people only post once in a while and only post things that are of significance (e.g., wedding pictures).

In such cases, an additional mechanism needs to be employed to ensure that the user is not flooded with notifications. One possible mechanism is to set a hard capacity constraint where once a user receives five notifications, no additional notification will be allowed. However, this would lead to efficiency loss as valuable notifications generated later on will no longer be pushed to users. Alternatively, we implement a soft constraint that dynamically updates prices: whenever a notification is sent to a user who has already received five notifications, their price is increased to the value that would have resulted in only five notifications being sent.

\subsection{Comparing Algorithm \ref{algo:online} and Production}

\begin{small}
\begin{table}[]
    \centering
    \begin{tabular}{l|c|c|c}
        & \textbf{FP with} & \textbf{FP with} & \textbf{SP with} \\
        \textbf{Metric} & \textbf{\footnotesize UB pacing} & \textbf{\footnotesize BB pacing} & \textbf{\footnotesize BB pacing} \\
        \hline
        average winning valuation & 0.5027 & 0.5113 & 0.4974 \\
        \hline
        number of notifications sent & 138,300 & 142,593 & 99,012 \\
        {\textit{(i). comment subscribed}} & 1,498 & 1,496 & 931 \\
        {\textit{(ii). feed suite org. campaign}} & 60,674 & 65,140 & 46,131 \\
        {\textit{(iii). like}} & 13,688 & 14,023 & 10,759 \\
        {\textit{(iv). story daily digest}} & 62,440 & 61,934 & 41,191 \\
        \hline
        supply constraint violation & & \\
        {\textit{(i). overall rate using $s_j$}} & 10.88\% & 10.32\% & 5.05\%  \\
        {\textit{(i). overall rate using $s_j \times 2$}} & 0.55\% & 0.32\% & 0.19 \% \\
        {\textit{(ii). average per user}} & 0.2216 & 0.1907 & 0.0872  \\
        \hline
        supply wastage & & \\ 
        {\textit{(i). overall rate}} & 77.72\% & 78.10\% & 84.25\% \\
        {\textit{(ii). average per user}} & 0.0780 & 0.1174 & 0.8324 \\
        \hline
        std.~dev.~of pacing multipliers & & \\
        {\textit{(i). comment subscribed}} & 0.39 & 0.27 &  0.94 \\
        {\textit{(ii). feed suite org. campaign}} & 0.22 & 0.81 & 245.78 \\
        {\textit{(iii). like}} & 1.57 & 0.74 & 10.93 \\
        {\textit{(iv). story daily digest}} & 1.49 & 1.86 & 727.71\\
    \end{tabular}
    \caption{Comparison of the outcomes of Algorithm \ref{algo:online} with heuristics in Section \ref{sec:heuristics} and budget spent based pacing. \vspace{-2em}}
    \label{tab:simulation-compare}
\end{table}
\end{small}

In this section, we investigate the performance of Algorithm \ref{algo:online} with the heuristics described in Section \ref{sec:heuristics}. In addition, using an internal simulator, we compare the outcome with the production pacing and pricing system (as described in Section \ref{sec: exp_sp_launch}).

The dataset we run the simulation on is obtained by sampling $0.01\%$ of the users during a ten-day period with the last three days used as the testing duration and the rest used for learning parameters. The exact window of data needed for learning stable parameters is tuned for each algorithm separately. The utility based pacing algorithm uses the three-day window just before the last three days to learn user prices and the budget based pacing algorithm uses the full seven-day window before the testing duration to learn stable system bidder values.

As is the case in the production implementation, we assume that the platform valuation $v^t_p = 0$ for all $t$. We summarize the results in Table \ref{tab:simulation-compare} and we next explain the metrics investigated.

The \emph{average winning valuation} is the average valuation for the notification type among all notifications that are sent to users. The \emph{number of notifications sent} is the number of notifications (of each notification type) that are sent to users. The \emph{standard deviations} of pacing multipliers measure the variability of the pacing multipliers for each notification type over time.

Next, for \emph{supply constraint violation}, \emph{overall rate} is the percentage of the users for whom the number of notifications sent is above the allowance. Here, we consider the original allowance $s_j$ as well as a softer allowance $2\times s_j$. The \emph{average per user} is the average number of notifications that are sent above the allowance $s_j$. 

Lastly, for \emph{supply wastage}, \emph{overall rate} is the percentage of the users for whom at least one rejected  notification can be sent without violating the supply constraint, and \emph{average per user} is the average number of notification opportunities that got wasted.

Immediately from Table \ref{tab:simulation-compare}, we can see that the outcomes are similar between the two FP auction systems with either pacing mechanisms, but they are quite different from the SP auction system. The main differences are (1) FP auction results in more notifications sent to users without sacrificing the quality of the sent notifications, and in fact, FP has a higher average winning valuation; (2) FP auction has a higher supply violation rate, although pretty small with the soft allowance $2 \times s_j$; (3) FP auction has lower wastage; and (4) pacing multipliers under FP auction have smaller variability, and thus FP auction has a more stable pacing system (see Figure \ref{fig:pacing-stability} in Appendix \ref{sec:app_exp}).

In their raw form, it is hard to evaluate whether sending more notifications to users is beneficial to the platform or not. However, in the next section, based on an FP versus SP production A/B test, we discuss further the effects of sending more notifications on other metrics of interest that are not directly observable from offline simulations.

\subsection{Online  A/B test: FP versus SP}

\begin{small}
\begin{table}[]
    \centering
    \begin{tabular}{l|c|c}
        & \textbf{SP with} & \textbf{FP with} \\
        \textbf{metric} & \textbf{\footnotesize BB pacing} & \textbf{\footnotesize BB pacing} \\
        \hline
        average winning valuation  & 0.6616 & 0.6610 \\
        \hline
        number of notifications sent & 8,064,73 & 8,314,501 \\
        {\textit{(i). comment subscribed}} & 26,657 & 30,681\\
        {\textit{(ii). feed suite org. campaign}}  & 2,140,6040 & 2,290,939 \\
        {\textit{(iii). like}}  & 167,433 & 212,628 \\
        {\textit{(iv). story daily digest}}  & 5,730,045 & 5,780,253 \\
        \hline
        supply constraint violation & & \\
        {\textit{(i). overall rate using $s_j$}} &  1.62\% & 1.46\%  \\
        {\textit{(i). overall rate using $s_j \times 2$}} & 0.05\% & 0.07\% \\
        {\textit{(ii). average per user}}& 0.03 & 0.03  \\
        \hline
        supply wastage & & \\ 
        {\textit{(i). overall rate}} & 95.21\% & 95.30\% \\
        {\textit{(ii). average per user}} & 0.52 & 0.48 \\
        \hline         
        std. dev. of pacing multipliers (no outliers) & &  \\
        {\textit{(i). comment subscribed}} & 55.21 (1.26) & 1.28 (1.08) \\
        {\textit{(ii). feed suite org. campaign}}  & 98.44 (0.92) & 0.40 (0.40) \\
        {\textit{(iii). like}}  & 7.32 (1.06) & 0.90 (0.88) \\
        {\textit{(iv). story daily digest}}  & 132 (2.03) & 1.12 (0.86) \\
    \end{tabular}
    \caption{Comparison between the outcome of first price and second price auctions, both with second price pacing, as measured in a three-day window of the A/B test. \vspace{-2em}}
    \label{tab:FP-SP-QE-1}
\end{table}
\end{small}

From a production implementation perspective, changing SP auction to FP auction is easy. But it is more involved to change the pacing system. Thus, given that our offline simulation shows similarity between FP auctions with different pacing systems, we ran an A/B test where we compared the FP auction system with the SP auction system, both with budget spent based pacing, which is currently used in production and the same as the simulator.

This A/B test was run between December 14, 2022 - Jan 19, 2023 on the same four push notification types at Instagram mentioned earlier. The test group (FP) and control group (SP) each consisted of 8.75 million users. 

The metrics from the A/B test are presented in Tables~\ref{tab:ab_test_results} and~\ref{tab:FP-SP-QE-1}. We observe that the test group sends more notification with smaller amount of supply constraint violation per user and supply wastage per user. But this increase does not lead to a change in the topline metrics such as click-through rate or any of the two previously mentioned core platform metrics. Thus, the additional notifications were likely targeted to users who are already active on the platform.

We observe that FP leads to a decrease in the variance of pacing multiplier in the test group for all the notification types. For two of the four notification types, there is at least an order of magnitude decrease in standard deviation. This reduction comes largely from reducing the magnitude of large spikes (Figure \ref{fig:pacing-stability-FP-QE} and Figure \ref{fig:pacing-stability-change-FP-QE} in Appendix \ref{sec:app_exp}). Note that spikes are possible in the notification system due to upstream production ecosystem changes or non-stationarity in the user activity on the platform. For example, during the early part of the experiment, a new machine learning model for assigning valuations to one of the notification types was launched in production. This launch impacted both the test and control groups. We observe that there are spikes in pacing multipliers for both the FP and SP auctions. But the spikes are much smaller in magnitude for FP. Further, the stability in FP pacing multipliers is visible not just in periods of change, but also other times. In Table \ref{tab:FP-SP-QE-1}, we observe that even when the outliers in pacing multipliers are clipped to the 5th or 95th percentile values, the standard deviation of FP is at least 10\% smaller than that of SP and for two notification types, it is more than 50\%  smaller.

A system with reduced variance has several desirable properties. Reduced variance means a more stable pacing system. Smaller fluctuations in pacing multiplier imply that the budget spending over time is closer to the ideal spending rate. Further, a stable pacing system would be easier to debug. Finally, if the pacing mechanism can graciously accommodate changes in upstream systems, then the process of launching changes in other parts of the ecosystem becomes smoother.

\begin{figure}[t]
    \centering
    \includegraphics[width=\linewidth]{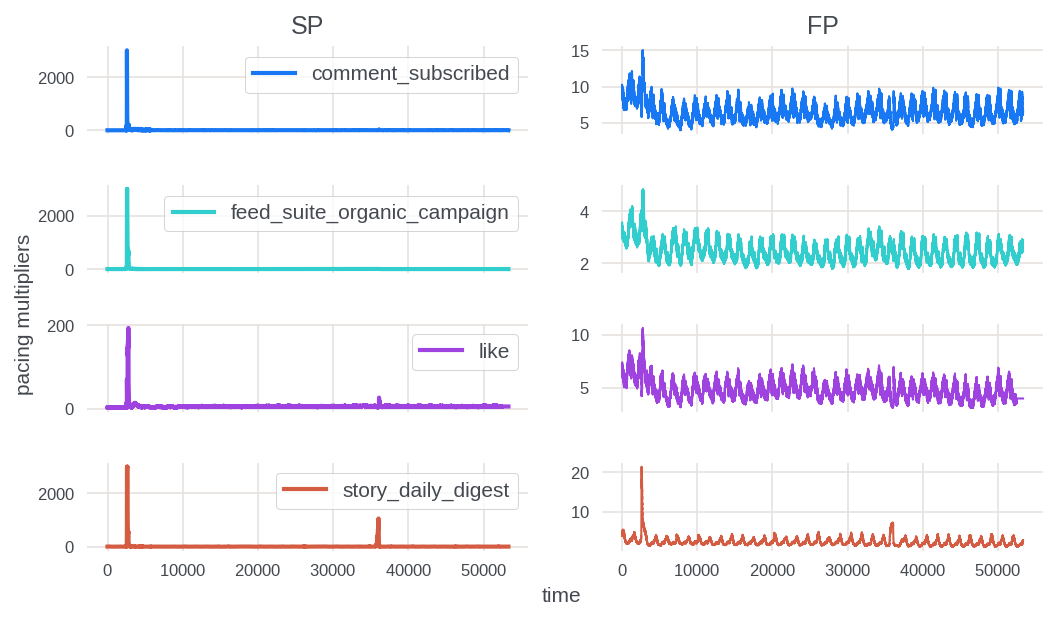}
    \caption{Stability of SP pacing multipliers (left) vs FP (right) in the A/B test with time (in minutes). Note the difference in the scale of the y-axis.}
    \label{fig:pacing-stability-FP-QE}
\end{figure}

\section{Conclusion And Future Work}

We presented an auction-based system for notification optimization. We provided theoretical guarantees for this auction system and illustrated its success in the production system for Instagram notifications.

There are several open questions related to this auction system, both from a theoretical and a system design perspective. First, as the system is expanded to include more notification types, we need a systematic approach to choose individual budgets (which are akin to priorities). The current system treats the notification delivery time decision and the notification sending decision as two separate problems (with our work focusing on the latter). It is worth investigating if combining these decisions into one framework can lead to further improvements. In particular, it is useful to understand the change in performance of the auction system if notification generation can be batched and several notifications simultaneously participate in the same auction instance.

When using Algorithm \ref{algo:online}, we need to estimate prices of the opportunities to send users notifications via historical data, and update the price when a user's capacity constraint is violated. The latter is a heuristic to deal with non-stationairity in user behaviors which results in spikes in the number of generated notifications for the user. It would be ideal to have a theoretically sound way of handling these spikes and, similar to what we have for the pacing multipliers, have a mechanism to learn these prices online.

\section*{Acknowledgements}
We would like to thank Bolun Zhang, Nailong Zhang, Moon Kang, Hao Feng, Haoran Zhang, and Zheqing (Bill) Zhu for their efforts in productionalizing the notification auction system.

\bibliographystyle{unsrtnat}
\bibliography{refs}

\appendix

\input{appendix.tex}

\end{document}

%% file: model_notation.tex
\section{Model and Notations}

We propose an economic approach to the notification prioritization problem. Here we treat this as a market equilibrium problem and solve it via an auctions-based approach.

To that end, we view the notification types as \emph{buyers} and denote the set of notification types by $T = \{\tau_1, \tau_2, \cdots, \tau_n\}$. The buyers are then interested in buying a set of \emph{items}, which are the individual opportunities to show a notification to a user. Thus, each user has a set of items associated to them. We denote the set of users by $U = \{u_1, u_2, \cdots, u_m\}$. Throughout the paper, we index notification types and users by $i\in [n]$ and $j\in [m]$ respectively. Each user $u_j$ has a capacity $s_j$ representing the maximum number of notifications that can be sent to the user within the time horizon $\horizon$. 
In order to enable our auctions-based approach, each notification type $\tau_i \in T$ is endowed with a budget of faux currency, where $B_i$ denotes the allocated budget. Since the budgets (and later prices) are in fake currency, the magnitude of the budget can be thought of as the level of priority a given notification type has. The larger the budget, the more notification opportunities they will be able to purchase relative to other notification types. 
By allocating items to buyers in the  economy induced from the faux currency, we can choose an allocation based on computing a \emph{competitive equilibrium}; in the case where all buyers have the same budget, this is called \emph{competitive equilibrium from equal incomes} (CEEI), which is known to have many attractive fairness properties~\citep{varian1974equity}.
Unlike the standard CEEI setup, we extend the model to include the platform as a buyer as well, and we let $B_p$ denote the platform's allocated budget.
The platform uses its budget to subsidize bids from notification types based on how beneficial a particular notification is to the platform.

We next describe the dynamics of the notification system. At a given time $t\in [\horizon]$, a notification of type $\tau_{i(t)}\in T$ is generated for user $u_{j(t)} \in U$, where we let $i(t)$ and $j(t)$ be the indices of the notification type and user respectively. We denote by $v^t>0$ (resp.~$v^t_p \geq 0$) the valuation of sending the generated notification to user $u_{j(t)}$ for the notification type $\tau_{i(t)}$ (resp.~for the platform). The valuation in this case can be interpreted as the probability of being clicked, or the probability of generating further interactions between users on the platform or some specific features of the platform. This valuation is typically aligned with the team's goals. Note that we allow notification types and platform to have different valuations. Moreover, in our setting, the valuations are time dependent, meaning that the same type of notification generated for the same user could have different valuations depending on other factors: for instance, different notifications informing users about comments on their posts can have varying click probabilities depending on who commented.

Once a notification is generated for a user, the platform then needs to decide whether the notification will be sent to the user. 
In our proposed auction-based system,
this decision relies on various pieces of information maintained by the system: 
\begin{enumerate} 
    \item[(i)] The \emph{price} $p_{j(t)}^t$ to send notifications to user $u_{j(t)}$ at time $t$; 
    \item[(ii)] A \emph{pacing multiplier} $\beta_{i(t)}^t$ of notification type $\tau_{i(t)}$ at time $t$; 
    \item[(iii)] A \emph{pacing multiplier} $\beta_p^t$ of the platform at time $t$. 
\end{enumerate} 
At time $t$, notification type $\tau_{i(t)}$ generates a \emph{bid} $\beta_{i(t)}^t v^t$, which is interpreted as their willingness to pay for the notification to be sent. Similarly, the platform generates a bid $\beta_p^t v^t_p$.
The generated notification is pushed to the user if and only if the sum of the bids from  the notification type $\tau_{i(t)}$ and the platform exceeds the price for user $j(t)$ at time $t$, i.e. if: 
\begin{equation}
    \underbrace{\beta_{i(t)}^t v^t}_{\textup{bid from notification type}} + \underbrace{\beta_p^t v^t_p}_{\textup{subsidy from platform}} \ge \underbrace{p_{j(t)}^t}_{\textup{price}}.
    \label{eq:allocation rule}
\end{equation}

If the notification is pushed to the user, then the notification type $\tau_{i(t)}$ and the platform gain utilities of $v^t$ and $v^t_p$ respectively.

The prices and pacing multipliers are maintained so that the \emph{utilities} of the notification types and the platform are maximized in a way that is \emph{fair} to all, while ensuring the number of notification sent to each user $u_j$ does not exceed $s_j$ (i.e., the supply constraint). To be more precise, let $x^t \in [0,1]$ be the probability with which the notification generated at time $t$ is sent to user $u_{j(t)}$, and for each notification type $\tau_i\in T$, let $\horizon_i \coloneqq \{t\in [\horizon]: i(t) = i\}$ be the collection of times during which a notification of type $\tau_i$ is generated. Then, the utility $u_i$ of each notification type $\tau_i$ and the utility $u_p$ of the platform are 
\begin{equation} \label{eq:utility}
    u_i\coloneqq \sum_{t\in \horizon_i} v^t x^t, \, \forall i\in [n]; \;\;\; u_p\coloneqq \sum_{t\in [\horizon]} v^t_p x^t.
\end{equation}

Now we turn to the question of how to select the prices.
It is known that such  auction setups with budgets can be captured by a \emph{Fisher market equilibrium}~\citep{borgs2007dynamics,conitzer2022pacing}. Our setting can be described as a novel type of Fisher market where there is an additional auxiliary buyer, the platform, whose utility depends on the overall allocation of items to buyers, whereas individual buyers (i.e. notification types) only care about their own allocation.

Next we define a competitive equilibrium for our setting.
For every notification type $\tau_i$, the bundle received by $\tau_i$ can be written as $x^{\horizon_i}$, which is a sub-vector of the allocation $x$. 
Let  $p \in \mathbb R^{\mathcal T}_+$ be a price vector for each possible notification.
Then we define the  \emph{demand set} $D_i(p,\beta_p)$ for notification type $\tau_i$ as follows,
$$\resizebox{.47\textwidth}{!}{$\displaystyle D_i(p,\beta_p) \coloneqq \arg \max_{x\in [0,1]^{\horizon_i}} \left\{\sum_{t\in \horizon_i} x^t v^t \mid \sum_{t\in \horizon_i} x^t (p_{j(t)}^t - \beta_p v^t_p) \le B_i \right\}$}.$$ 
Similarly, define the demand set of the platform as 
$$\resizebox{.47\textwidth}{!}{$\displaystyle D_p(p,\beta) \coloneqq \arg \max_{x\in [0,1]^{[T]}} \left\{\sum_{t\in [\horizon]} x^t v^t_p \mid \sum_{t\in [\horizon]} x^t (p_{j(t)}^t - \beta_{i(t)} v^t) \le B_p \right\}$}.$$ 

For each user $u_j\in U$, we similarly let $\horizon_j \coloneqq \{t\in [\horizon]: j(t) = j\}$ as the collection of times during which a notification is generated for user $u_j$.
We define a \emph{competitive equilibrium} to be a tuple $(x,p,\beta)$ of allocation $x$, price vector $p \in \mathbb R_+^{\mathcal T}$, and pacing multipliers $\beta\in \mathbb R^n_+,\beta_p\in \mathbb R_+$ such that:
\begin{enumerate}
    \item (notification types get demands) Notification types receive bundles from their demand sets: $x^{\horizon_i} \in D_i(p,\beta_p), \; \forall \tau_i\in T$. \label{cond demand}
    \item (platform gets demand) Platform receives a bundle from its demand set: $x\in D_p(p,\beta)$. \label{cond platform demand}
    \item (supply) For every user $u_j$, $\sum_{t\in \horizon_j} x^t = \min(s_j, |\horizon_j|)$. \label{cond supply}
\end{enumerate}

Note that we defined prices as being on a per-notification basis. Alternatively, one could try to define prices as being per \emph{user}, meaning that each user $u_j$ gets a price $p_j \geq 0$, and the price for a particular notification at time $t$ would be $p_{j(t)}$. We will use such per-user pricing in the second-price setting, but for the first-price setting we will need per-notification pricing.

In the typical Fisher market equilibrium, the equality condition in the supply constraint is usually only set as a less-than-or-equals, with equality required only if the price for a user is strictly positive. Since we will have per-notification pricing, it is simpler to avoid dealing with whether the \emph{user} price is positive and thus we have \Cref{cond supply} in the definition of competitive equilibrium.

We define a \emph{notification-type competitive equilibrium} (NCE) to be a tuple $(x,p,\beta)$ such that conditions (1) and (3) are satisfied, but the platform may not receive a bundle in its demand set. We will see that first-price auctions will allow us to obtain competitive equilibria, whereas second-price auctions will only allow NCE.

%% file: first-price-auctions.tex
\section{First-Price (FP) Auction} \label{sec:FP}

In this section we describe our proposed approach for how to utilize \emph{first-price auctions} to determine which notifications to send. Our first-price auctions will still use the rule in \cref{eq:allocation rule} to decide whether to send or not. If we decide to send, then everyone pays their bid, meaning that the notification type pays $\beta_{i(t)}^t v^t$ and the platform pays $\beta_p^t v^t_p$.
We will show that a competitive equilibrium can be used to facilitate such first-price-based allocation for our problem.
In order to develop this equilibrium notion, we will start by considering a \emph{static} setting where we know the full set of notifications that will be generated, and thus we do not have to worry about the online aspect of allocation. \cref{sec:algorithms} will then develop online algorithms for handling the online arrival of notification generation.

It is known that a competitive equilibrium in the standard Fisher market model can be found by finding an allocation that maximizes the \emph{budget-weighted geometric mean} of the utilities for the buyers.  For our setting with an auxiliary platform buyer, the budget-weighted geometric mean is
$\left(  u_p^{B_p} \times \prod_{i\in [n]} u_i^{B_i}  \right)^{\frac{1}{B_p + \sum_{i\in [n]} B_i}}.$
Now, in order to obtain a convex program for finding such a maximizing allocation, one can take the logarithm of the geometric mean and drop the resulting constant $\frac{1}{B_p + \sum{i\in [n]} B_i}$ to get an Eisenberg-Gale (EG) \cite{eisenberg1959consensus}-style convex program: 
\begin{small}
\begin{align} \label{eq:mnw}
\begin{split}
    \text{max}_x \quad & \sum_{i\in [n]} B_i \log(u_i(x)) + B_p \log(u_p(x)) \\
    \text{s.t.} \quad & \sum_{t\in \horizon_j} x^t \le s_j, \; \forall j\in [m] \\
    & x^t \in [0,1], \; \forall t\in [\horizon].
\end{split}
\end{align}
\end{small}

For every user $u_j$, let $p_j$ be the dual variable associated with its corresponding supply constraint. This dual variable can be interpreted as the ``base price'' of sending a notification to user $u_j$. For each time $t$, let $\lambda^t$ be the dual variable associated with the $x^t\le 1$ constraint and define the per-notification price $p_{j(t)}^t\coloneqq p_{j(t)}+\lambda^t$. Moreover, let $\beta_i = B_i / u_i$ and $\beta_p = B_p/ u_p$ be the pacing multipliers of buyers and the platform.
First-order optimality conditions and complementary slackness yield that for all $t\in [\horizon]$, we have:
\begin{equation}\label{eq:foc}    
\begin{aligned} 
    \beta_{i(t)} v^t + \beta_p v^t_p \leq p_{j(t)}^t = p_{j(t)} &\text{ if } x^t = 0 \\
    \beta_{i(t)} v^t + \beta_p v^t_p = p_{j(t)}^t = p_{j(t)} &\text{ if } x^t \in (0,1) \\
    \beta_{i(t)} v^t + \beta_p v^t_p = p_{j(t)}^t \geq p_{j(t)} &\text{ if } x^t =1.
\end{aligned}
\end{equation}
Thus, we can interpret this as notification type  $\tau_{i(t)}$ facing a clearing price of $p_{j(t)} - \beta_p v_p^t$, with the platform subsidizing the amount of $\beta_p v_p^t$. 
The allocation occurs only if the notification type bids beats the clearing price.
Since we run a first-price auction, the total payment by the notification type and the platform is $p_{j(t)}^t$ instead of the base price $p_{j(t)}$, out of which the notification type $\tau_{i(t)}$ pays $\beta_{i(t)} v^t$ and the platform pays $\beta_p v_p^t$

By definition of the pacing multipliers and the inequalities given in \eqref{eq:foc}, we immediately have the following two propositions (proofs of all propositions and theorems are presented in Appendix \ref{sec:app_proofs}) . 

\begin{proposition}[budget clearing]
    \label{prop:budget-spent}
    For each notification type $\tau_i$ and the platform, we have 
    $$\sum_{t\in \horizon_i} \beta_i v^t x^t = B_i, \; \forall i\in [n]; \;\; \textup{ and } \sum_{t\in [\horizon]}  \beta_p v_p^t x^t = B_p.$$
\end{proposition}

\begin{proposition}[budget spent optimally] \label{prop:opt-spend}
    For each notification type $\tau_i$, and any $t_1, t_2 \in [\horizon]$ such that $i(t_1)=i(t_2)=i$, if $x^{t_1} > 0$, then 
    $$\frac{p_{j(t_1)}^{t_1}-\beta_p v^{t_1}_p}{v^{t_1}} \le \frac{p_{j(t_2)}^{t_2}-\beta_p v^{t_2}_p}{v^{t_2}}.$$ 
    Similarly, for the platform, let $t_1, t_2\in [\horizon]$, if $x^{t_1}> 0$, then 
    $$\frac{p_{j(t_1)}^{t_1}-\beta_{i(t_1)} v^{t_1}}{v^{t_1}_p} \le \frac{p_{j(t_2)}^{t_2}-\beta_{i(t_2)} v^{t_2}}{v^{t_2}_p}.$$ 
\end{proposition}

It now follows that \cref{eq:mnw} yields a competitive equilibrium; we call a competitive equilibrium from \cref{eq:mnw} a \emph{notifications first-price pacing equilibrium} (nFPPE).
\begin{theorem} \label{thm:CE-FP}
    An optimal solution $x$ for \eqref{eq:mnw}, the prices $\{p_{j(t)}^t\}_t$ defined based on dual variables, and the pacing multipliers $\beta_i = B_i/u_i, \forall i\in[n]$ of buyers and $\beta_p = B_p/u_p$ of the platform yield a competitive equilibrium.
\end{theorem}

Another consequence of \cref{eq:mnw} is that an nFPPE must necessarily be Pareto optimal: if there exists some alternative allocation that Pareto improves the utilities, then it would yield a strictly higher objective value in \cref{eq:mnw}, which contradicts optimality.

\input{proportionality.tex}

%% file: proportionality.tex
\smallskip
\textbf{Fairness Guarantees for nFPPE.} We now show that nFPPE satisfies two new fairness criteria that mirror standard fairness criteria in fair allocation but adapted to our setting.
Assume without loss of generality that $|\horizon_i\cap\horizon_j| \geq s_j$ for every notification type $\tau_i$ and every user $u_j$. 
This is wlog because if this does not hold then we can add zero-valued opportunities until it does.

Consider the allocation $\tilde x$ where each notification type $\tau_i$ is given a fraction $f_i\coloneqq B_i / (B_p+\sum_{i\in[n]}B_i)$ of the total supply $s_j$ of each user, and then notification type $\tau_i$ consumes $s_j f_i / |\horizon_i\cap\horizon_j|$ of each opportunity $t\in \horizon_i\cap\horizon_j$.
Let $\underline u_i = \sum_{t\in\horizon_i} v^t \tilde x^t$ be the utility of this allocation.
Consider also the allocation $\hat x$ where the platform gets its proportional share: $\hat x^t = (s_{j(t)} / \horizon_{j(t)}) \times (B_p/(B_p+\sum_{i\in[n]}B_i))$ for all $t$. 
Let $\underline u_p$ be the platform utility of this allocation $\hat x$.
We say that an allocation $x$ satisfies \emph{proportionality} if for each notification type $\tau_i$, $\sum_{t\in\horizon_i} x^tv^t \geq \underline u_i$, and for the platform, $\sum_{t\in[\horizon]} x^t v_p^t \geq \underline u_p$.
\begin{proposition}
    Any nFPPE allocation $x$ satisfies proportionality.
    \label{prop:nFPPE-proportionality}
\end{proposition}

%% file: second-price-auctions.tex
\section{Second-Price (SP) Auctions} \label{sec:SP}

Next we study a model where we use second-price auctions to perform each allocation. 
This will still use the allocation rule in \cref{eq:allocation rule}, but the difference is in the price charges to a winning notification type. 

At a given time $t$, suppose that the notification type bid plus the platform subsidy is greater than $p^t_{j(t)}$. In that case, we charge the notification type $p_{j(t)} - \beta_p^tv_p^t$, which is exactly the lowest amount that they could have bid in order to get allocated.
Note that the platform pays its full subsidy, which is $\beta_p^tv_p^t$.

We now study the offline version of this model. In that case, our goal will be to find what we will call a \emph{notifications second-price pacing equilibrium} (nSPPE).
An nSPPE is a tuple $(x, p, \beta)$ where $x$ is an allocation, $p\in \mathbb R_+^m$ is a set of prices for the users, and $\beta=(\{\beta_i\}_{i\in [n]}, \beta_p) \in \mathbb R_+^{n+1}$ is the set of pacing multipliers for notification types and the platform, such that:
\begin{itemize}
    \item[(1)] (price setting) The price $p_j$ for user $u_j$ equals to the $(s_j+1)$-th highest combined bid (i.e., notification type bid plus platform subsidy) for notifications generated for $u_j$, or zero if there are no more than $s_j$ notifications generated for $u_j$. 
    \item[(2)] (winning bids allocated) $x^t = 1$ if $\beta_{i(t)}v^t + \beta_p^t v_p^t \ge p_{j(t)}$ and $x^t=0$ otherwise.
    \item[(3)] (supply exhausted) For every user $u_j$, $\sum_{t\in \horizon_j} x^t \leq s_j$ where equality holds if $p_j > 0$.
    \item[(4)] (notification type budgets cleared) For every notification type $\tau_i$: $\sum_{t\in \horizon_i} x^t (p_{j(t)} - \beta_p v_p^t) = B_i$. 
    \item[(5)] (platform budget cleared)  $\sum_{t\in \horizon} x^t  \beta_p v_p^t = B_p$. 
\end{itemize}

SPPE was previously introduced by \citet{conitzer2022multiplicative} in the setting of second-price auction markets with budgets, where each buyer is quasilinear, meaning that they have value for leftover budget. Here we consider a variant of their SPPE concept where buyers (i.e. notification types) do not have outside value for their leftover budget. Moreover, our model is a generalization that allows for the existence of a platform bidder and multiple items per user (i.e. the notification opportunities).

\begin{proposition} \label{prop:opt-spend-sp}
    In an nSPPE, each notification type $\tau_i$ buys a  bundle in their demand set given the prices.
\end{proposition}


It follows that an nSPPE is a competitive equilibrium among the notification types given the subsidized prices, i.e. an NCE. However, the platform does not buy a bundle that maximizes its utilities with the prices defined as the per-notification-opportunity price minus the notification-type bid, since that is not even the price faced by the platform in nSPPE.

\smallskip
\textbf{Fairness Guarantees for nSPPE.} 
It turns out that nSPPE guarantees a certain notion of proportionality.
Consider the allocation $\bar x$ where every notification type receives their budget-proportional share (i.e., $f_i$) of the supply of each user, and then allocates that supply optimally to the notification opportunities they have for that user. We call the utilities of notification types under $\bar x$ their \emph{proportional share utility}. Proposition below states that in an nSPPE, the utility of each notification type is at least as good as their proportional share utility.
Note that this guarantee is stronger than the proportionality guarantee for nFPPE, where the buyers are not allocated towards their highest-value items for each user. %

\begin{proposition} \label{prop:proportion-notif-type-sp}
In an nSPPE $(\beta, p, x)$, each notification type receives an allocation such that $\sum_{t\in \horizon_i} x^t v^t \geq \sum_{t\in \horizon_i} \bar{x}^t v^t$.
\end{proposition}

Unfortunately, the nSPPE solution concept is unlikely to be easily computable, unlike our first-price solution concept which can be solved via convex programming due to \cref{eq:mnw}. It was recently shown that for the SPPE solution concept of \citet{conitzer2022multiplicative}, it is PPAD-hard to compute an SPPE~\citep{chen2021complexity}, which makes it unlikely that an efficient algorithm for nSPPE exist.
Despite the theoretical drawback on computability, nSPPE has an important practical advantage over nFPPE: existing pacing systems at large-scale internet companies are targeted at pacing in second-price auction markets (despite that problem falling directly under the PPAD-hardness result of \citet{chen2021complexity}) and thus nSPPE can be implemented relatively easily by leveraging such existing infrastructure.

\iflong

\ck{add existence result?}

\fi

%% file: appendix.tex
\section{Additional Tables}
\label{sec:app_tables}

Table \ref{tab:notif_desc} gives the description of notification types and their budgets. The budgets are scaled by a certain factor to hide their magnitude.
\begin{small}
\begin{table}[ht]
    \centering
    \begin{tabular}{p{2.4cm}|p{4.0cm}|p{1cm}}
        \bf Notification Type & \bf Description & \bf Budget (scaled) \\
        \hline
        Comment Subscribed & Informs about comments on  posts on which the user is subscribed & 100 \\
        \hline
        Feed Suite Organic Campaign & Informs users about new posts on their feed & 3500 \\
        \hline
        Like & Informs about ``like'' reaction(s) on a post by the user & 2887 \\
        \hline
        Story Daily Digest &  Informs about other users who recently added an Instagram story & 4448
    \end{tabular}
    \caption{Description and budgets of notification types \vspace{-1.5em}}
    \label{tab:notif_desc}
\end{table}
\end{small}

\section{Additional Proofs}
\label{sec:app_proofs}
\subsection{For Section \ref{sec:FP}}

\begin{proof}[Proof of Proposition \ref{prop:budget-spent}]
    The first claim follows because for each notification type $\tau_i$, we have $\sum_{t\in \horizon_i} \beta_i v^t x^t = (B_i/u_i) \sum_{t\in \horizon_i}  v^t x^t = (B_i/u_i) u_i = B_i$. The same logic holds for the platform.
\end{proof}

\begin{proof}[Proof of Proposition \ref{prop:opt-spend}]
    This can be shown by rearranging the terms in the first-order conditions in \eqref{eq:foc}.
\end{proof}

\begin{proof}[Proof of Theorem \ref{thm:CE-FP}]
    Items \ref{cond demand} and \ref{cond platform demand} from our definition of a competitive equilibrium follow from \cref{prop:opt-spend,prop:budget-spent}. \cref{cond supply} follows because if the claim is not true, then there must be one buyer who has a $t$ such that they could improve their utility by getting more of $x^t$, which contradicts \Cref{cond demand}. 
\end{proof}

\begin{proof}[Proof of Proposition \ref{prop:nFPPE-proportionality}]
We know 
from \cref{prop:budget-spent} that
\begin{small}
\begin{align}
    B_p + \sum_{i\in[n]} B_i 
    =&\sum_{t\in \horizon} p_{j(t)}^t x^t \nonumber\\
    \geq&\sum_{t\in \horizon_i, x^t=1} p_{j(t)}^t  x^t + \sum_{t\notin \horizon_i} p_{j(t)} x^t + \sum_{t\in \horizon_i, x^t < 1} p_{j(t)} x^t \label{eq:prop price}
\end{align}
\end{small}
Now consider any bundle $\hat x$ purchased by notification type $\tau_i$ such that $\sum_{t\in \horizon_j \cap \horizon_i} \hat x^t = s_j$ for each user $u_j$.
Notice that the price faced by $\tau_i$ for buying $\hat x$ is upper bounded by \cref{eq:prop price}, since if $x^t = 1$ then $x^t \geq \hat x^t$, and thus the price paid for taking the whole supply $s_j$ of user $u_j$ is $\sum_{t\in \horizon_j \cap \horizon_i, x^t = 1} \hat x^t p_{j}^t +  \sum_{t\in \horizon_j \cap \horizon_i, x^t < 1} \hat x^t p_{j}$, which is weakly less than the total price contributed by user $u_j$ in \cref{eq:prop price}.

Thus, we have that the allocation $\hat x$ costs at most $B_p + \sum_{i\in[n]} B_i$. It follows directly that $\tilde x$ costs at most $B_i$, and thus notification type $\tau_i$ could have afforded the proportional bundle $\tilde x$. By \cref{prop:opt-spend}, $\tau_i$ must prefer their bundle in $x$ to buying $\tilde x$.

By a completely analogous argument, proportionality holds for the platform as well.
\end{proof}

\subsection{For Section \ref{sec:SP}}

\begin{proof}[Proof of Proposition \ref{prop:opt-spend-sp}]
    First note that every notification type spends their budget by definition of nSPPE.
    Secondly, consider any $t\in \horizon_i$. If $x^t > 0$, then we have that the price per utility is at most $\beta_i$ because the notification type bid $\beta_{i} v^t$ is at least $(p_{j(t)} - \beta_p^t v_p^t)$; on the other hand if $x^t=0$, it must be that $\beta_i v^t < (p_{j(t)} - \beta_p^t v_p^t)$. Thus, goods they win yield a better return on spend (i.e., best bang per buck) than goods they do not win. Moreover, for any $t$ such that $\beta_i > (p_j - \beta_p^t v_p^t) / v^t$, we have by definition that $i$ wins all of that good, and thus they buy as much as possible of any good with price-per-utility strictly greater than $\beta_i$.
\end{proof}

\begin{proof}[Proof of Proposition \ref{prop:proportion-notif-type-sp}]
    It follows from the definition of nSPPE that $\sum_j s_j p_j = B_p + \sum_{i\in [n]}B_i$. 
    Thus the total price of the allocation $\bar x$ is $B_p + \sum_{i\in [n]}B_i$, since it also exactly allocates $s_j$ of each user $u_j$. Since notification type $\tau_i$ receives $B_i / (B_p + \sum_{i\in [n]}B_i)$ of each user, their share $\bar x^{\horizon_i}$ costs exactly $B_i$. 
    Therefore they can afford their proportional bundle (which can only become cheaper due to platform subsidies), and since they are buying a utility-maximizing bundle given the prices, it must be the case that they receive at least as much utility.
\end{proof}

\subsection{For Section \ref{sec:algorithms}}

\begin{proof}[Proof of Proposition \ref{prop:dual}]
    We can rewrite the problem in \eqref{eq:mnw} by adding constraints $u_i\le \sum_{t\in \horizon_i} x^t v^t$ for all $i\in [n]$ and $ u_p \leq \sum_{t\in \horizon} x^t_p v^t_p$. Let $(p_j)_j$, $(\lambda^t)_t$, $(\beta_i)_i$, and $\beta_p$ be the Lagrange multipliers on the corresponding supply constraint, the $x^t\le 1$ constraint, the new utility constraints on notification types and the platform, respectively. Then, we can apply Sion's minimax theorem (this applies because we can add redundant upper bounds to the allocation variables $x_{ij} \leq s_j+\epsilon$ without loss of generality), we get the Lagrangian of the optimization problem:
    \begin{align*}
        & \min_{p, \lambda, \beta \geq 0} \; \max_{x, u\geq 0} \; \sum_{i=1}^n B_i \log u_i + B_p \log u_p - \sum_{j=1}^m p_j(\sum_{t\in \horizon_j} x^t -s_j) \\
        & - \sum_{t=1}^{\horizon} \lambda^t (x^t-1) - \sum_{i=1}^n \beta_i(u_i - \sum_{t\in \horizon_i} x^t v^t) - \beta_p(u_p - \sum_{t\in \horizon} x^t_p v^t_p) \\
        =& \min_{p, \lambda, \beta \geq 0} 
        \sum_{i=1}^n \max_{u_i \geq 0} (B_i \log u_i - \beta_i u_i)
        + \max_{u_p \geq 0} (B_p \log u_p - \beta_p u_p ) \\
        & + \sum_{j=1}^m p_j s_j + \sum_{t=1}^{\horizon} \lambda^t - \sum_{t=1}^{\horizon} \max_{x \ge 0} x^t (p_{j(t)} + \lambda^t - \beta_{i(t)} v^t - \beta_p v_p^t).
    \end{align*}
    Here we note the fact that for strictly positive constants $B,\beta$, we have that $\max_u B\log u - \beta u = B\log B - B\log \beta - B$. In addition, while $\beta_i\geq 0$ is the constraint in the Lagrangian, we must have $\beta_i > 0$ because otherwise, $\max_{u_i} B_i\log u_i - \beta_i u_i = \max_{u_i} B_i\log u_i = +\infty$ (this is handled by the $\log\beta$ term in the derivation). Similarly, we must have $\beta_p>0$. Solving each of the single-dimensional optimization problems, we get that the Lagrangian is equivalent to
    \begin{align*}
        \min_{p\geq 0,\beta \geq 0} 
        & \sum_{i=1}^n (-B_i \log \beta_i + B_i\log B_i - B_i) \\ 
        &+ (-B_p \log \beta_p + B_p\log B_p - B_p)
        + \sum_{j=1}^m p_j s_j + \sum_{t=1}^{\horizon} \lambda^t \\
        \textup{s.t.}\;\; & p_{j(t)} \geq \beta_{i(t)} v^t + \beta_p v_p^t -\lambda^t, \; \; \forall t\in [\horizon].
    \end{align*}
    After removing constants from the objective, we get \cref{eq:mnw dual}.
\end{proof}

\begin{proof}[Proof of Theorem \ref{thm:pace}]
    This follows by interpreting PACE as dual averaging (DA) in the composite setting without an auxiliary regularizer. 
    Consider the problem
    \[
        \min_w \E f_z(w) + \Psi(w),
    \]
    where $f_z(w)$ is a convex and subdifferentiably function for each point $z\in Z$ for some arbitrary sample space, and $\Psi$ is a strongly convex function. 
    At iteration $t$, let $f_t$ be the observed convex function based on some sample $z$, let $g^t\in \partial f_t(w^t)$, and then the \emph{dual average} is $\bar g^t = \sum_{\tau=1}^t \frac{1}{t} g^t$.
    DA without an auxiliary regularizer uses the update rule
    \[
        w^{t+1} = \arg\min_w \langle \bar g^t,w\rangle + \Psi(w).
    \]
    Going back to \cref{eq:mnw pairs inf}, we let $f_z = \max_i (\beta_i v_{it} + \beta_p v_{pit} - p_j)$ (as a function of $\beta$), and $\Psi = -\sum_i B_i\log \beta_i - B_p\log\beta_p$, with domain equal to $\beta_i \in [0, B_i / \underline u_i]$ for each $i\in \{1,\ldots,n,p\}$. Now it is straightforward to see that $u_i^t$ is the subgradient $g^t_i$, $\bar u_i^t$ is the dual average, and the update rule for pacing multipliers is the DA rule.
    Finally, by \cref{prop:nFPPE-proportionality} we have that $\beta_i^* = B_i / u_i^* \in [0, B_i / \underline u_i]$.

    Now, invoking \citet[Theorem 2]{gao2021online}, which is exactly the theorem statement when $G^2$ is an upper bound on the squared $\ell_2$ norm of the subgradients and $\sigma^2$ is the strong convexity parameter, yields the bound. Here, note that $\sigma^2$ is indeed the strong convexity parameter, which can be seen by calculating the second derivative of $-B_i\log \beta_i$ which is $B_i / \beta_i^2$, and noting that the corresponding Hessian has diagonal entries. 
    So, a lower bound on the Eigenvalues is a lower bound on $\min_i B_i / \beta_i^2$ over all $i$ including the platform. But we know $\beta_i \leq B_i / \underline u_i$, so we get $B_i / \beta_i^2 \geq B_i / (B_i / \underline u_i) = \underline u_i$.
\end{proof}

\section{Additional Figures and  Experiment Results}
\label{sec:app_exp}

\begin{figure}[ht]
    \centering
    \begin{subfigure}[t]{.48\linewidth}
    \centering
    \includegraphics[width=\textwidth]{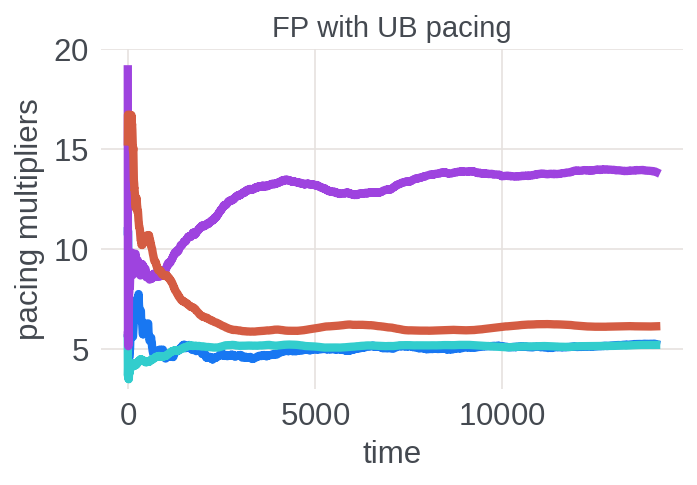}
    \end{subfigure}%
    \begin{subfigure}[t]{.48\linewidth}
    \centering
    \includegraphics[width=\textwidth]{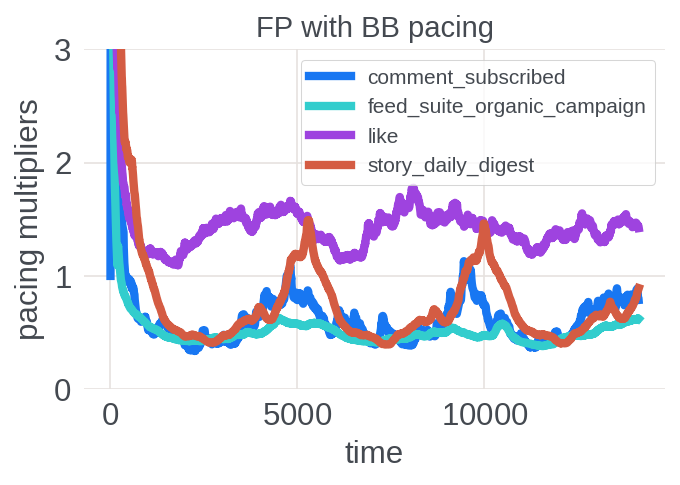}
    \end{subfigure}\\
    \begin{subfigure}[t]{.48\linewidth}
    \centering
    \includegraphics[width=\textwidth]{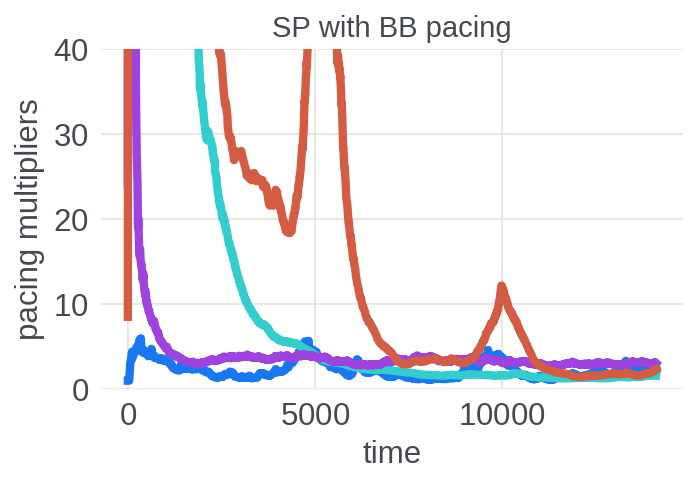}
    \end{subfigure}%
    \begin{subfigure}[t]{.48\linewidth}
    \centering
    \includegraphics[width=\textwidth]{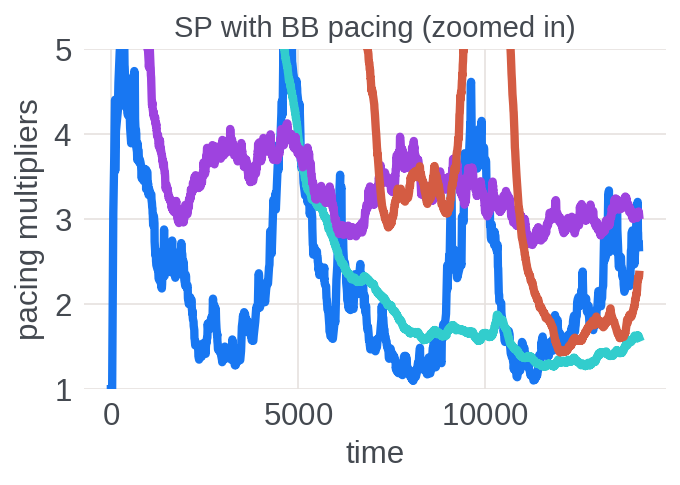}
    \end{subfigure}%
    \caption{Pacing multipliers over time (in minutes), comparing three settings investigated in our offline simulation. Note that pacing multipliers are more stable under first price auctions. \vspace{-.5em} }
    \label{fig:pacing-stability}
\end{figure}

\begin{figure}[ht]
    \centering
    \includegraphics[width=\linewidth]{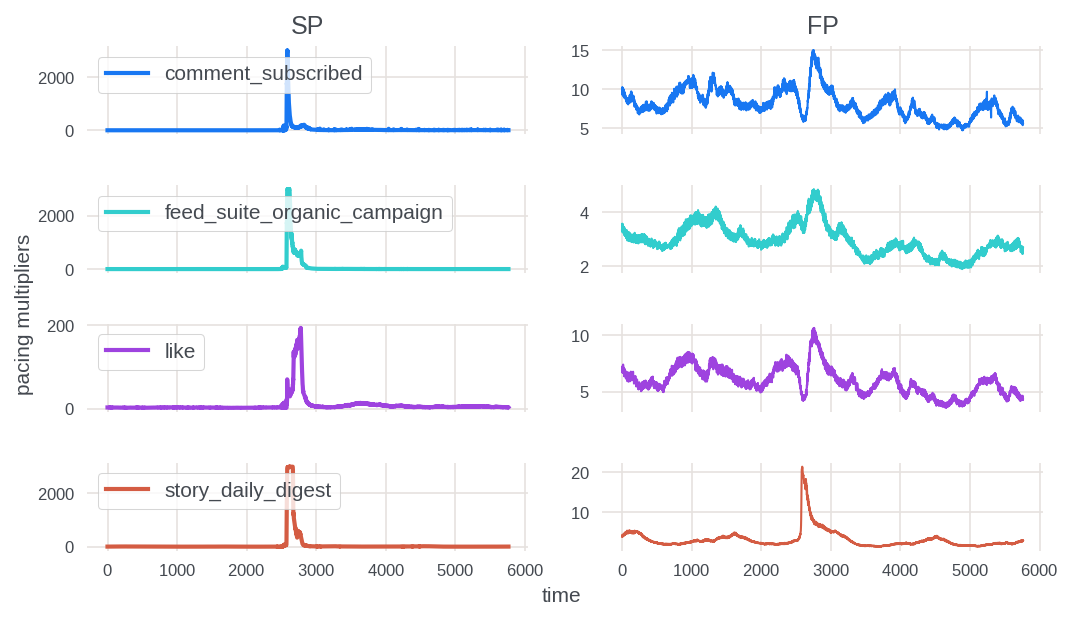}
    \caption{During a period of change in an upstream system, this figure shows that pacing multipliers have much smaller spikes with FP auction as compared to SP auction in the A/B test}
    \label{fig:pacing-stability-change-FP-QE}
\end{figure}